\documentclass[amssymb,useAMS,prd,aps,amsmath,nofootinbib,superscriptaddress,twocolumn]{revtex4}                
\usepackage{graphicx}
\usepackage{amssymb}
\usepackage{color}
\usepackage{multirow}
\usepackage{amsmath}
\usepackage{feyn}
\usepackage{feynmf}
\usepackage{hyperref}
\usepackage{wasysym}
\usepackage[caption=false]{subfig}
\usepackage{ulem}
{

\begin{document}
%\title{Searching for GeV-scale Dark Gauge Bosons in QGP thermal di-%lepton production}
\title{Searching for GeV-scale new Gauge Bosons in QGP thermal dilepton production}
\author{Jonathan H. Davis}
\affiliation{Institute for Particle Physics Phenomenology, Durham University, Durham, DH1 3LE, United Kingdom}
\author{C\'eline B\oe hm}
\affiliation{Institute for Particle Physics Phenomenology, Durham University, Durham, DH1 3LE, United Kingdom}
\affiliation{LAPTH, U. de Savoie, CNRS,  BP 110,
  74941 Annecy-Le-Vieux, France}
\date{\today}

\begin{abstract}
In this paper we propose to use the measurement of the thermal  Quark-Gluon Plasma (QGP) di-lepton spectra in the Intermediate Mass Region (IMR) of heavy-ion collisions, as a new method to search for GeV-scale dark gauge bosons ($\gamma'$ or $Z'$). Such light mediators are a common feature of light (i.e. low mass)  dark matter scenarios, which have been invoked to explain puzzling signals in dark matter indirect and direct detection experiments. First we show that a light $\gamma'$ or $Z'$ will generate a resonant enhancement of the di-lepton spectrum produced thermally by the QGP, at an energy corresponding to the dark gauge boson mass. Secondly, using data from the PHENIX experiment, we are able to set an upper limit on the combined coupling of this new gauge boson to quarks and leptons   (independently of their vectorial or axial nature)  $ \chi_q \chi_e < 10^{-3}$ at the $95 \% $ confidence level for a  gauge boson mass $m \in [1.5,2.5] \, \mathrm{GeV}$. This result complements previous searches for new light gauge bosons and probes a new region of the parameter space, particularly interesting in the case of non-universal couplings to quarks and leptons. Prospects for the discovery of such a boson by the ALICE collaboration are also discussed. 
\end{abstract}

\maketitle

\section{Introduction}
\label{sec:intro}
%==================================

With the recent discovery of the Higgs boson \cite{Aad:2012tfa,Chatrchyan:2012ufa}, the validity of the Standard Model (SM) is no longer under question for energies up to the electroweak scale. This remarkable achievement, together with  the lack of evidence for physics beyond the SM at the LHC and in other Particle Physics experiments, might suggest that there exist no other particles than those of the SM. However the unknown nature of dark matter and the origin of neutrino masses still provide two convincing indications that the SM needs to be extended. 
 
The current `Vanilla'  paradigm assumes that Dark Matter is made of very heavy particles, with a mass in the GeV-TeV range. No obvious sign of dark matter particles has been found as yet above 10 GeV. However there exist a few puzzling (although still controversial) signals at low energy e.g. \cite{Jean:2003ci,Boehm:2003bt,Bernabei:2010mq,Aalseth:2010vx,Angloher:2011uu,Agnese:2013rvf} which may suggest that dark matter is relatively light.
Meanwhile new phenomenological directions have started to emerge during the last decade and new types of dark matter candidates have been proposed. In particular there has been an increasing interest for light dark matter candidates interacting with new (light) gauge bosons. These scenarios were first proposed in \cite{Boehm:2002yz,Boehm:2003hm}  based on cosmological arguments but they are now advocated to reconcile contradicting results from recent dark matter direct detection experiments (see for example  \cite{Frandsen:2011cg,Frandsen:2013cna,Schwetz:2011xm,Chang:2010yk}).

One natural scale for both the dark matter and gauge boson in these frameworks is of a few GeVs \footnote{Such a scale can be motivated by theories beyond the Standard Model \cite{Fayet:1990wx,Fayet:2007ua,Abel:2008ai}.}. Hence if these particles exist, they should be accessible in low energy Particle Physics experiments, especially those with a good integrated luminosity. But more than their mass range, the estimate of the strength of their couplings to quarks and leptons is of direct relevance to the experimental set-up. 

A number of constraints have already been placed in the literature on new (spin-1) gauge boson couplings. Generally one assumes either purely vectorial (in which case the dark boson is referred to as $\gamma'$ or dark photon) or vectorial and axial ($Z'$) couplings.  Heavy dark boson couplings to quarks have been constrained in \cite{Carena:2004xs,Jaeckel:2012yz,Aad:2012hf,Chatrchyan:2012oaa}, assuming a mass $\gtrsim 50 \, \mathrm{GeV}$.  Light (sub-GeV) dark photons coupling to quarks have also been constrained using hadronic decay channels (e.g. $\phi \rightarrow e^+ e^-$ \cite{Babusci:2012cr,Reece:2009un}, $\eta$ and $\eta^{\prime}$ decays \cite{Gninenko:2012eq}, Kaon decays \cite{Beranek:2012ey} and $J/\psi$ decays \cite{PhysRevD.75.115004}).  Additional limits on the quark and lepton couplings were set from parity-violation experiments \cite{Bouchiat:2004sp,Fayet:2007ua} (on the relative size of the axial and vector couplings, in the case of a $Z'$ boson) and, in the case of gauge bosons lighter than $\lesssim 1 \, \mathrm{GeV}$, from neutrino experiments \cite{Boehm:2004uq,Chiang:2012ww}, beam dump as well as fixed-target experiments \cite{PhysRevD.86.095019,Abrahamyan:2011gv,Bjorken:1988as,PhysRevD.80.075018}. 

However the GeV-10 GeV range remains relatively unconstrained. At present the most relevant limit in this mass range  has been set using data from the BaBar experiment \cite{Reece:2009un,PhysRevD.80.075018,PhysRevD.80.015003,Aubert:2009au,Echenard:2012iq}. Assuming universal couplings to all leptons, the ratio of the dark photon-lepton coupling to the ordinary photon-lepton was constrained to be $\chi_e \sim 2\cdot10^{-3}$ for $m_{\gamma'} \in [0.5,10] \, \mathrm{GeV}$. However at present no robust bound on the coupling to quarks has been set yet.

Here we develop a novel idea. We propose to look for a new resonance in the dilepton  spectrum associated with heavy ion collisions, in order to search for light (GeV) gauge bosons, relevant to DM scenarios.  Dilepton signals are modern tracers of the formation of a Quark-Gluon plasma (QGP) in heavy-ion collisions and have been studied in detail by the PHENIX collaboration \cite{Adare:2009qk}, and more recently at the ALICE experiment at CERN \cite{Koehler:2013mew}, for both proton-proton and heavy ion collisions. By investigating the presence (or lack) of a resonance in the dilepton spectrum, from heavy-ion collisions, in the Intermediate Mass Range with respect to the theoretical predictions, we show that it is possible to obtain meaningful constraints on new GeV gauge bosons coupled to both quarks and leptons (and possibly to the dark matter).

Note that we will focus on the contribution from thermal partonic production in the QGP, and neglect prompt collisions (e.g. Drell-Yan from partons in the colliding nuclei), which are significantly weaker than the thermal emission in the IMR (see section \ref{sec:Zprime}).

In Section \ref{sec:QGP}, we discuss the present status of dilepton production in the Quark-Gluon plasma. In Section \ref{sec:Zprime}, we determine the signature of new gauge bosons 
in QGP experiments such as PHENIX  and derive constraints on the gauge boson couplings. We discuss possible improvement on this limit in   Section \ref{sec:future} and conclude in 
Section \ref{sec:conclusion}.

\section{Quark-Gluon plasma}
\label{sec:QGP}
%==================================

The formation of a QGP in high energy heavy-ion collisions has been debated for decades, however recent experimental data have confirmed its existence. A simple picture of the QGP is as a thermal gas of de-confined quarks and gluons, formed in the early stages of high-energy heavy-ion collisions due to the large QCD energy densities present. Under such conditions a phase transition, or possibly a crossover, occurs, where the partons are no longer bound into hadrons or mesons, and remain so until the energy density (or temperature) drops below some critical value. This is characterised in lattice simulations as a rapid increase in the number of relevant degrees of freedom, as the temperature of the matter produced in nuclear collisions rises above this critical value \cite{PhysRevD.79.074505,Gyulassy:2004zy}.

In what follows we first discuss the evidence and theoretical efforts to model the QGP formation and dilepton signals.

%\vspace{-21pt}
\subsection{Experimental evidence}

A strong indication for QGP formation in  heavy-ion collisions is an excess of dileptons over the predicted contributions from hadronic decays and Drell-Yan production, for an invariant mass $m_{ee}$ of GeV-scale \cite{Adare:2009qk,Arnaldi:2009,Abreu:2000,PhysRevC.63.054907,Linnyk:2011vx}.  Multiple theoretical explanations have been proposed as to the origin of this excess: an enhanced contribution from decays of $c$ and $\bar c$ quarks was successful in fitting early data \cite{Abreu:2000}. However with more data \cite{Adare:2009qk,Arnaldi:2009} such a model was disfavoured (evidence actually indicates a reduced $c \bar c$ contribution for nuclear collisions \cite{Adare:2009qk,PhysRevC.87.014905}), and was replaced instead with the far more successful scenario of dileptons originating from partonic interactions in a quark-gluon plasma (QGP), formed in nuclear-collisions.  

Although observations of  a dilepton excess provide compelling evidence for the formation of a QGP in heavy-ion collisions, such an emission could originate from another unknown source or enhanced background. However the observed suppression of high-energy hadrons in nuclear collisions with respect to proton-proton collisions \cite{Adcox:2001jp,Aamodt:2010jd,Chatrchyan:2011sx} provides additional arguments in favour the QGP scenario. The latter has a natural explanation in terms of the hadrons' transit through a strongly-interacting medium (supposed to be the quark-gluon plasma) causing them to lose energy through collisions and stimulated gluon emission.

Given such evidence, we will proceed to analyse the production of dileptons by the QGP in more detail, with the ultimate aim of fitting it to experimental data from the PHENIX experiment \cite{Adare:2009qk}.

\subsection{Modeling}
To determine the signature of light dark bosons, we first need a reliable estimate of dilepton production in heavy ion collisions. In the GeV energy range,  it is possible to use a perturbative treatment\footnote{The transition to perturbativity is not well-defined. However we will assume that perturbativity is valid at the energies considered in this paper.} to model the quark and gluon interactions responsible in the QGP for dilepton production \cite{Blaizot:2011ks,Blaizot:2010cc}. However since the plasma exists at finite temperature the perturbative series itself must be modified to account for its existence.

 For this purpose, it is convenient to consider the plasma constituents as quark and gluon partons with non zero thermal masses (in the perturbative regime) \cite{Braaten:1989mz}\footnote{This resummation also results in the modification of the quark-gluon vertex for soft momenta ($\sim gT$). This could potentially affect the $q + g \rightarrow q + e^+ e^-$ and $q + \bar q \rightarrow g + e^+ e^-$ processes, but should have only a small effect here since we work in the regime where the dilepton pair mass $m_{ee} > T$.}. These thermal masses regulate singularities in the amplitudes of photon production processes \cite{Blaizot:2005mj,Linnyk:2010vb} and are also required to improve the agreement with the findings from lattice field theory \cite{Blaizot:2005mj}. They scale with the temperature as $m_q \approx g T$ \cite{PhysRevC.87.014905,Kajantie:1982nj,Peshier:1995ty,Dusling:2009ej,Thoma:1999nm}, where $T$ is the QGP temperature and $g = \sqrt{4 \pi \alpha_s}$, the strong-interaction coupling.

In this work we will adopt the relation $m_q =  \sqrt{C_f}  g T / 2$, where $\alpha_s = 0.4$ and $C_f = (N_c^2 - 1)/(2 N_c)$, with $N_c = 3$, the number of colours \cite{Blaizot:2005mj}. For gluons we take $m_g = \sqrt{\frac{4}{3} \pi \alpha_s (N_c + N_f/2)} T$ \cite{Arleo:2004gn}, with $N_f = 3$, the number of light quark flavours (u, d, s) in the QGP. We will also model the dilepton excess observed in heavy ion collisions using perturbative thermal theory.

\subsection{Possible caveats}
 Such a resummation for obtaining thermal masses may not be enough to guarantee the accuracy of a perturbative approach, since it effectively treats the thermal partons as collision-less \cite{PhysRevD.62.096012}.  A full treatment of dilepton production would require the inclusion of processes due to scattering effects in the plasma, both through multiple scattering \cite{1126-6708-2002-12-006,PhysRevD.62.096012,PhysRevD.54.5274} and processes where the quark single-scatters then annihilates \cite{Aurenche:2002pc}. Multiple scattering (via gluon exchange)  occurs when the effective length for a quark to travel before emitting a low-invariant mass photon is larger than the mean free path in the plasma.  In the non-thermal theory the diagrams for such scattering processes would appear at higher-order in the perturbative expansion, but in the plasma each extra thermal quark propagator can effectively decrease the order of a diagram by $m_q^{-2} \propto \alpha_s^{-1}$ in the collinear regime \cite{1126-6708-2002-12-006}. These are generally referred to as ladder diagrams  \cite{Blaizot:2005mj}, representing an infinite series of scattering via gluon exchange inside a quark loop, and must be further resummed for a collisional medium such as the QGP \cite{1126-6708-2002-12-006,PhysRevD.54.5274}. In this case the scatterings can not be treated independently and will interfere with each other, which is a manifestation of the Landau-Pomeranchuk-Migdal (LPM) effect \cite{PhysRevD.62.096012}. Furthermore the effect of giving the quarks and gluons a finite width, due to multiple scattering interactions, is also considered in \cite{PhysRevD.62.096012,Linnyk:2010vb}. There are also tree-level contributions from the decays of thermal quarks and gluons, with the latter only possible in the plasma due to the gluon thermal mass \cite{Linnyk:2010vb}.

In each case the effects of such additional processes are at their largest when the virtual photon is approximately light-like, which corresponds to the low invariant mass regime \cite{1126-6708-2002-12-006,Gelis:2002yw} (in particular for the direct pair annihilation of $q \bar q$). In addition lattice results indicate that the weakly-coupled perturbative model of thermal partons works reasonably well at energy scales roughly at least several times larger than the QGP critical temperature $T_c \approx 170 \, \mathrm{MeV}$ \cite{Blaizot:2010cc,Blaizot:2005mj}.  As an example, a lattice simulation performed in \cite{PhysRevD.79.074505} determined the fluctuations in baryon number, strangeness and charge of the QGP. At energies a few times that of $T_c$ such fluctuations came only in packets consistent with a gas of free quarks (e.g. charge fluctuated only in units of the bare quark charge), indicating only weak modifications to the quarks behaviour from that of a collision-less gas.

Hence we restrict our analysis to the region where the dilepton invariant mass $m_{ee}$ is larger than the QGP temperature (specifically the region $1.2 \, \mathrm{GeV} < m_{ee} < 2.6 \, \mathrm{GeV}$) and consider the simplest case of a plasma of thermal partons, since contributions from non-perturbative effects should be sub-dominant. To compute the contributions from the multiple-scattering processes and resummation effects mentioned above, we use a publicly-available code \cite{Geliscode} but we do not compute such corrections for  the dark gauge boson. Note however that this does not mean it is exempt from LPM effects; it is possible that such processes (and for example the ISR of a $\gamma^{\prime}$ or $Z^{\prime}$) could have interesting effects beyond a simple resonance, perhaps even affecting dilepton emission at lower $m_{ee}$.

\subsection{Dilepton production for $1.2 \, \mathrm{GeV} < m_{ee} < 2.6 \, \mathrm{GeV}$}
\label{sec:dilepton_prod}

At  GeV-scale, the QGP is expected to be an abundant source of  dileptons  \cite{Adare:2009qk,Peitzmann:2001mz,Linnyk:2010vb,Linnyk:2011vx,Dusling:2009ej,Peshier:1994zf,Peshier:1995ty,PhysRevC.87.014905,Dusling2009212,PhysRevC.63.054907}, owing to  the exchange of a virtual photon in $q + \bar q \rightarrow e^+ e^-$, $q + g \rightarrow q + e^+ e^-$ and $q + \bar q \rightarrow g + e^+ e^-$ processes \cite{Dusling:2009ej,Linnyk:2010vb,Cleymans:1991ps}.

To obtain the full thermal dilepton spectra we will integrate over the phase-space and (simplified) space-time evolution of the plasma, assuming the quarks and gluons to be thermally distributed \cite{Peitzmann:2001mz}.  For quarks we take the Fermi-Dirac distribution ($f_{\mathrm{FD}}$) and for gluons that of Bose-Einstein ($f_{\mathrm{BE}}$). Before performing the space-time integration, the expression for dilepton production takes the form,
\begin{equation}
\frac{\mathrm{d}N}{\mathrm{d}^4 x} = \prod_i \left[ \int \frac{\mathrm{d}^3 p_i  f_{\mathrm{th}}(E_i)}{(2 \pi)^3 2 E_i} \right] |\mathcal{M}|^2 (2 \pi)^4 \delta^4 \left({\sum_j} P_j \right)
\label{eqn:phase_space_1}
\end{equation}
where $|\mathcal{M}|^2$ is the amplitude, $i$ runs over the participating particles with four-momentum $P_i = (E_i, p_i)$ and $f_{\mathrm{th}}(E) = f_{\mathrm{FD/BE}}(E)$ for initial-state coloured particles or $f_{\mathrm{th}}(E) = 1 \pm  f_{\mathrm{FD/BE}}(E)$ for final-state coloured particles, with $+$ for bosons and $-$ for fermions.

For simplicity one can assume that the QGP is in thermal and chemical equilibrium, in which case the chemical potential $\mu$ can be set to zero, and the densities of quarks and gluons are effectively equal. However this is likely to be too simplistic an assumption, as the QGP is expected to reach equilibrium only towards the end of its lifetime \cite{Gelis:2004ep}. In the initial stages of its out-of-equilibrium evolution one expects the QGP to be gluon-dominated \cite{d'Enterria:2005vz,Gelis:2004ep}, which can be represented by different values of $\mu$ for quarks and gluons, which change also as the plasma evolves. As a result, in this early phase the processes $q + g \rightarrow q + e^+ e^-$ is enhanced relative to $q + \bar q \rightarrow g + e^+ e^-$  and  $q + \bar q \rightarrow e^+ e^-$. We shall model this using temperature-dependent fugacities ($\lambda$) (see \cite{Gelis:2004ep}, however there exist alternative models e.g. \cite{PhysRevC.63.054907}), leading to a modified out-of-equilibrium distribution $f(E)$ of the form,
\begin{equation}
f_{\mathrm{non-eq}}(E) = \frac{\lambda_{q,g}(T)}{e^{E/T} \pm \lambda_{q,g}(T)}.
\end{equation}
As one can see the equilibrium is restored when $\lambda = 1$, bearing in mind that chemical potential and fugacity are related by $\mu = \mu_0 + k_B T \ln \lambda$, with $\mu_0 = 0$ in our case. Additionally the fugacity itself can be temperature-dependent and be different for quarks and gluons. Note that the thermal quark and gluon masses are modified slightly in the non-equilibrium case \cite{Arleo:2004gn}.

To account for the space-time evolution of the plasma, we  integrate from its initial creation, from which it cools from a temperature $T_{\mathrm{max}}$ to the critical temperature $T_0 = 170 \, \mathrm{MeV}$. We define $\mathrm{d}^4 x = V(\tau) \mathrm{d} \tau$, where for the volume $V$ and temperature $T$ of the plasma we use the Bjorken model \cite{PhysRevD.27.140}. This takes the plasma as forming in the region between two relativistic nuclei just after the collision; the high energy-density in this region allows the formation of coloured partons, which quickly thermalise through collisions. The expansion of this thermal plasma is longitudinal and homogeneous, hence we have \cite{Peitzmann:2001mz},
\begin{eqnarray}
V &=& 2 \pi R_N^2 \tau \\
T &\propto& \tau^{-1/3}.
\end{eqnarray}
The expressions are parameterised in terms of the plasma evolution time $\tau$, $R_N$ is the nuclear radius and $T(\tau = 0.2 \, \mathrm{fm}) = T_{\mathrm{max}}$.

In order to calculate the dilepton spectrum as a function of invariant-mass $m_{ee}$ we integrate Eqn.~\ref{eqn:phase_space_1}  (after integrating over $\mathrm{d}^4 x$) in discrete-bins of $m_{ee}$ and divide by the bin-size to get the average. We  take a bin-size of $\Delta m_{ee} = 0.25 \, \mathrm{GeV}$, to facilitate the  comparison with experimental data. Note that there is some subtlety involved in this calculation. First we integrate over the time $\tau$ in the inertial frame of the plasma itself \cite{PhysRevD.27.140}, while we seek to determine the dilepton spectrum in the lab frame.

These frames may actually differ due to the potential bulk motion of the plasma as it expands from the collision point. However since the dilepton spectrum is Lorentz-invariant our calculation should not be affected by any plasma bulk motion. There may be nevertheless some issues with cuts in pseudo-rapidity and $p_T$ in the data, since the cuts themselves are frame-dependent. This will likely affect the overall normalisation of the signal, which we discuss later.
%\vspace{-10pt}
\begin{figure}[h]
%\centering
\includegraphics[scale=0.45]{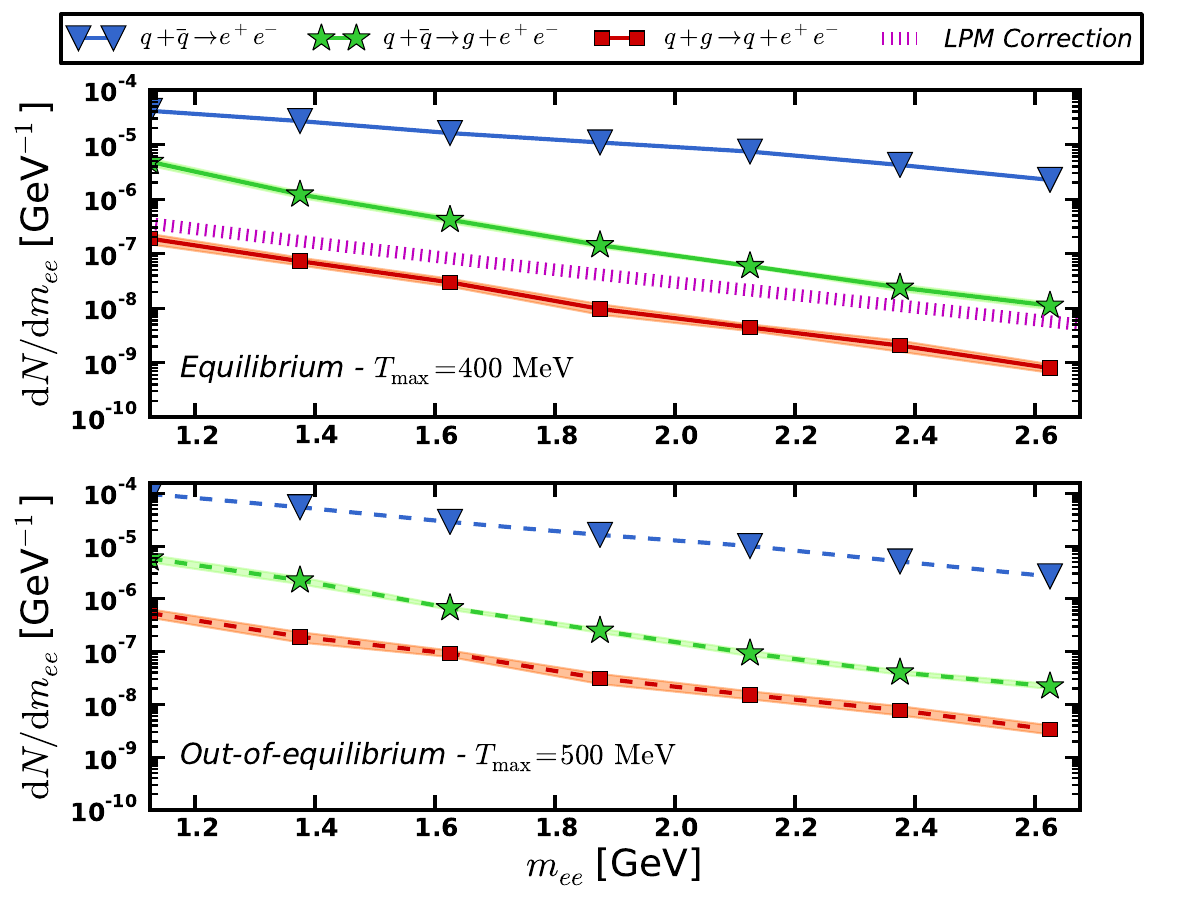}
\caption{Invariant mass spectra of dileptons produced thermally by various processes in the QGP, with initial temperature labelled as $T_{\mathrm{max}}$. The spectra in the top panel have been calculated assuming a plasma in equilibrium (i.e. equal fugacity for quarks and gluons $\lambda = 1$), while the lower panel takes the fugacities of quarks and gluons to be different \cite{Gelis:2004ep}, and gluon-dominated during the initial stages of evolution. The shaded bands indicate uncertainty in the Monte Carlo integration.}
\label{fig:mee_spectra_photon}
\end{figure}

The dilepton spectra for the processes discussed above are shown in fig. \ref{fig:mee_spectra_photon}.  A common  feature to these spectra  is the exponential drop with larger $m_{ee}$ \cite{Dusling2009212,Dusling:2009ej,Linnyk:2011vx,PhysRevC.87.014905,Peressounko:1999fs} for $m_{ee} \gtrsim 1 \, \mathrm{GeV}$. 
As one can see from this figure, the process $q \bar q \rightarrow e^+ e^-$ is the dominant mechanism of dilepton production for $m_{ee} \gtrsim 1 \, \mathrm{GeV}$, which is in agreement with other calculations of the dilepton spectrum in the IMR \cite{Dusling2009212,Dusling:2009ej,PhysRevC.87.014905,Linnyk:2011vx}. 

As expected, in the case of a non-equilibrium plasma both processes with initial state $q \bar q$ are suppressed relative to $q + g \rightarrow q + e^+ e^-$. Since the plasma is only strongly gluon-dominated during its initial stages, such an enhancement of the $qg$ process is not enough to make it competitive with the $q \bar q \rightarrow e^+ e^-$ process in the invariant-mass range considered here. Note also that the out-of-equilibrium plasma is expected to be slightly hotter \cite{d'Enterria:2005vz}, hence the overall rate from all three partonic processes is largely unchanged. Finally we find that the contribution from multiple-scattering, i.e. the Landau-Pomeranchuk-Migdal resummation (LPM) for dilepton production, is size-able, but remains nevertheless sub-dominant in the IMR.

Here we have taken the strong-coupling constant to be temperature-independent and fixed at $\alpha_s = 0.4$.  Finally another point to consider is the initial temperature of the plasma. The latter has a strong effect on the overall rate \cite{Peitzmann:2001mz,Peressounko:1999fs}. For RHIC a reasonable estimate of the initial temperature\footnote{There is ambiguity in this value, with several models for photon/dilepton production using different values in an approximate range from $300 \, \mathrm{MeV} \mbox{-} 600 \, \mathrm{MeV}$ \cite{Dahms:2008bs}. } (and the value we use for our analysis) is $T_{\mathrm{max}} = 400 \, \mathrm{MeV}$, assuming that nuclei collide at a centrality of $0\% \mbox{-} 20 \%$ \cite{Dusling:2009ej}.

This results in a photon spectrum of the same magnitude as previous calculations \cite{Dusling:2009ej,Linnyk:2011vx}. However comparing our result with that of \cite{Dusling:2009ej} (and as shown also in \cite{Adare:2009qk}) we see that our spectrum, although having a similar $m_{ee}$ dependence, is larger overall. The reason for this discrepancy is not known, however it could be due to the use of a different hydrodynamical model, or perhaps a different initial value of $\tau$ (which we take as $0.2 \, \mathrm{fm}$). We will proceed to use our calculated spectrum, however the impact on our results of altering the overall size, to match that of  \cite{Dusling:2009ej}, will be discussed in sec. \ref{sec:constraints}.

%\vspace{-10pt}
\section{Searches for new gauge bosons at PHENIX }
\label{sec:Zprime}

Since our calculations successfully reproduce previous determinations of the expected thermal QGP dilepton spectra, we can now study the contribution of a new virtual gauge boson to these spectra and confront our results to the Au-Au data from the PHENIX experiment \cite{Adare:2009qk}.

%\vspace{-10pt}
\subsection{New gauge boson characteristics}
The simplest implementation of a dark photon is to consider a  new (massive) particle with vector-like interactions, proportional to that of the photon (see \cite{Jaeckel:2013ija} for a review).  The ratio of the $\gamma^{\prime}$ coupling to that of the photon is labelled as $\chi_i$, with $i$ any SM particle that is electromagnetically charged. 
We thus have the following relation $Q^{\prime}_{i} = \chi_i Q_{i}$, where $Q_{i}$ is the charge of the SM particle $i$). Alternatively one can consider a gauge boson with possibly both vectorial and axial-vector couplings to quarks and leptons, like a $Z$ boson. Such a particle is generally referred to as a $Z^{\prime}$ and can have a different mass $m_{Z^{\prime}}$ and also suppressed couplings to the Standard Model particles, relative to the $Z$ (also labelled $\chi_i$). For simplicity hereafter we will assume a universal suppression for all quark flavours, but one can easily extend our results to non universal couplings.

\begin{figure}[h]
\centering
\subfloat{ \includegraphics[scale=0.13]{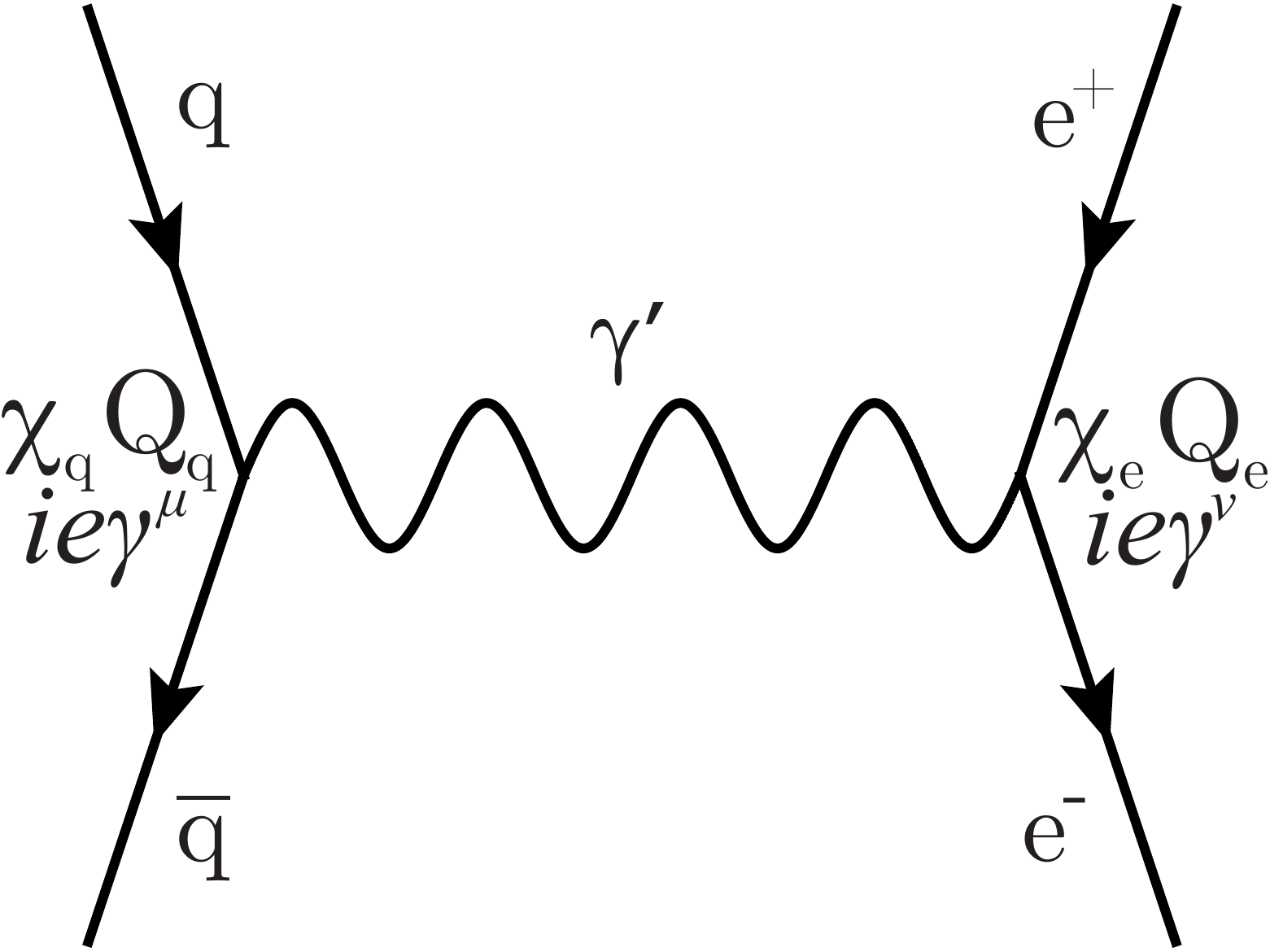}} \vspace{5pt}
\subfloat{ \includegraphics[scale=0.13]{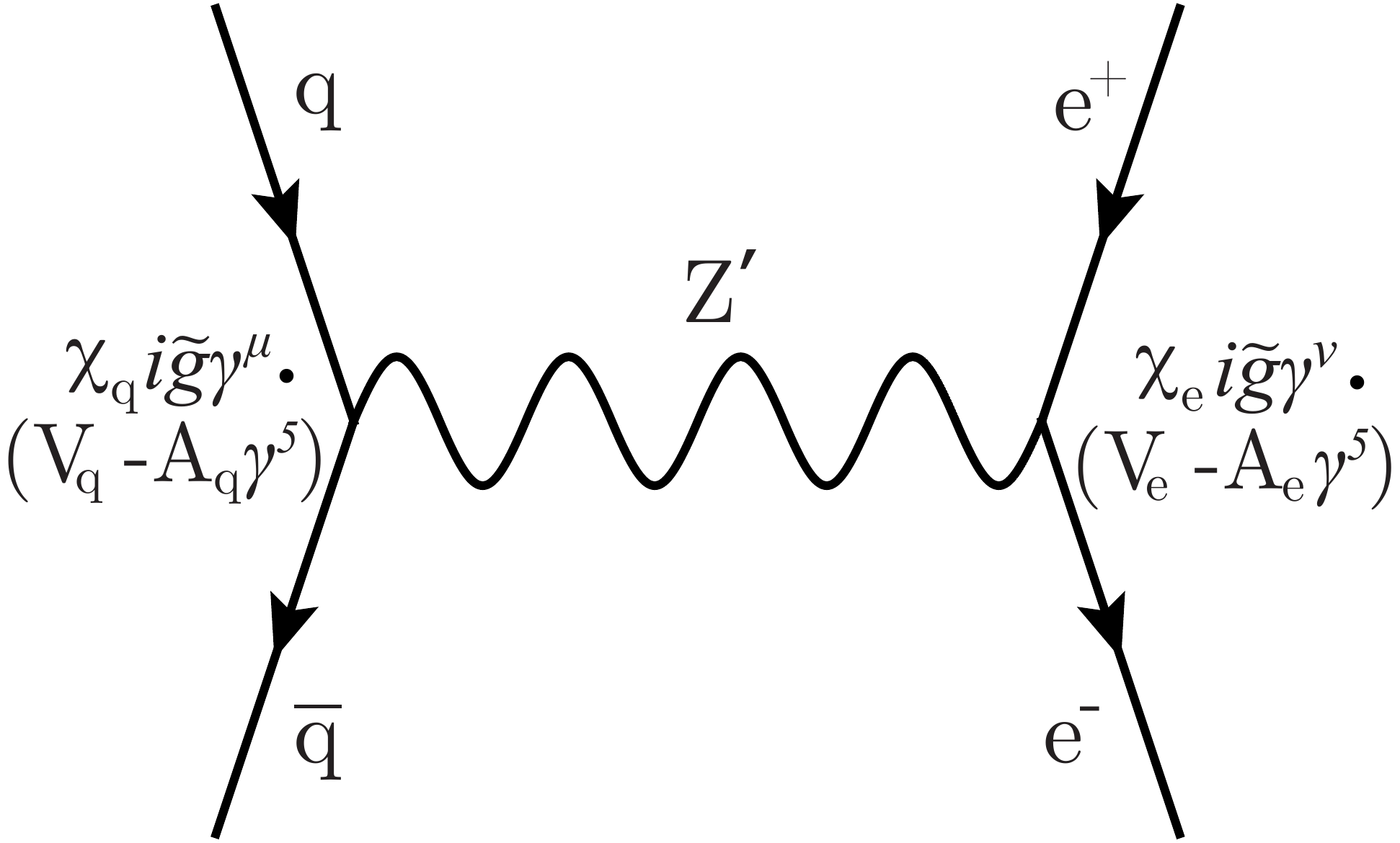}} \\
\subfloat{ \includegraphics[scale=0.13]{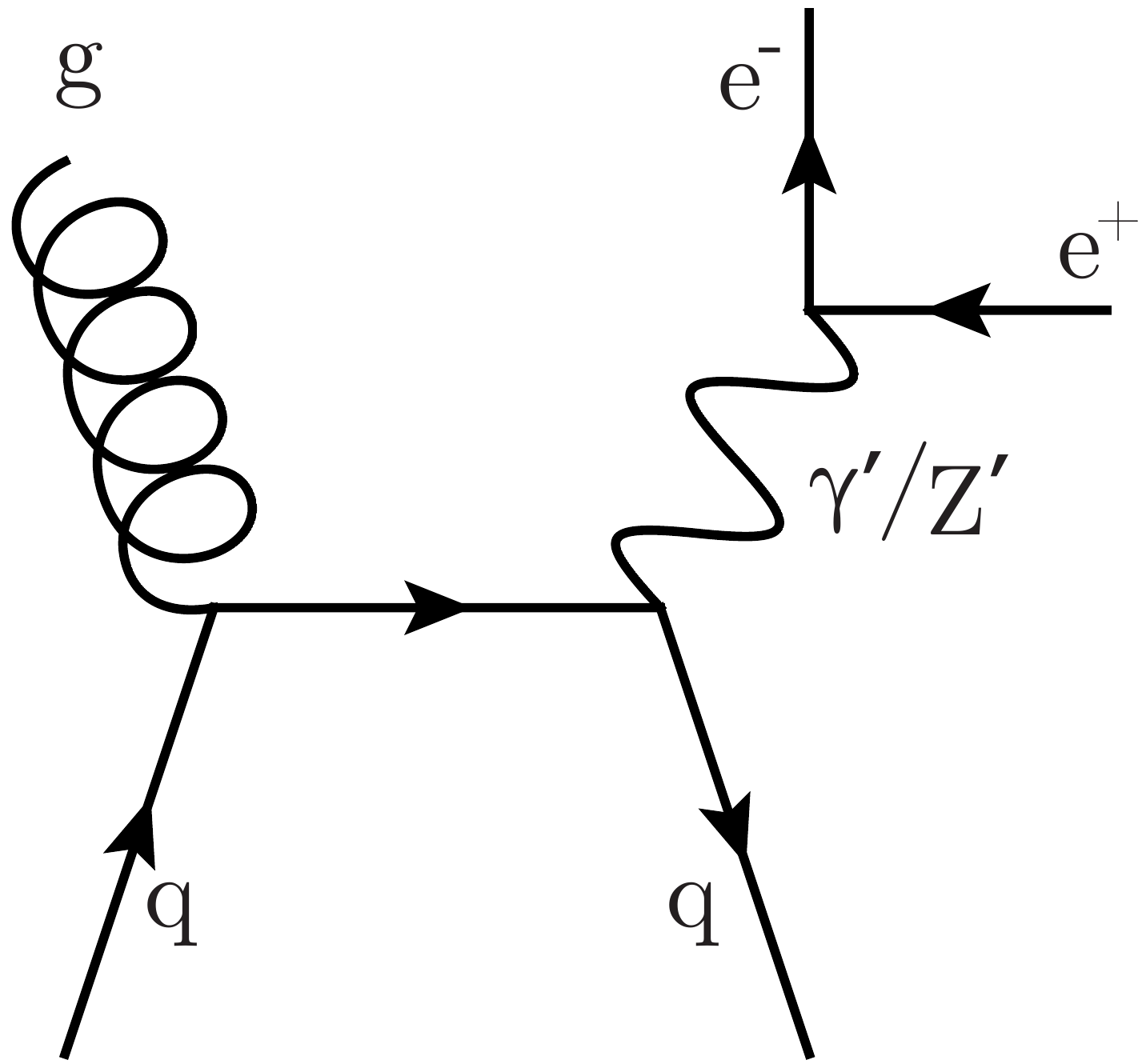}}
\subfloat{ \includegraphics[scale=0.13]{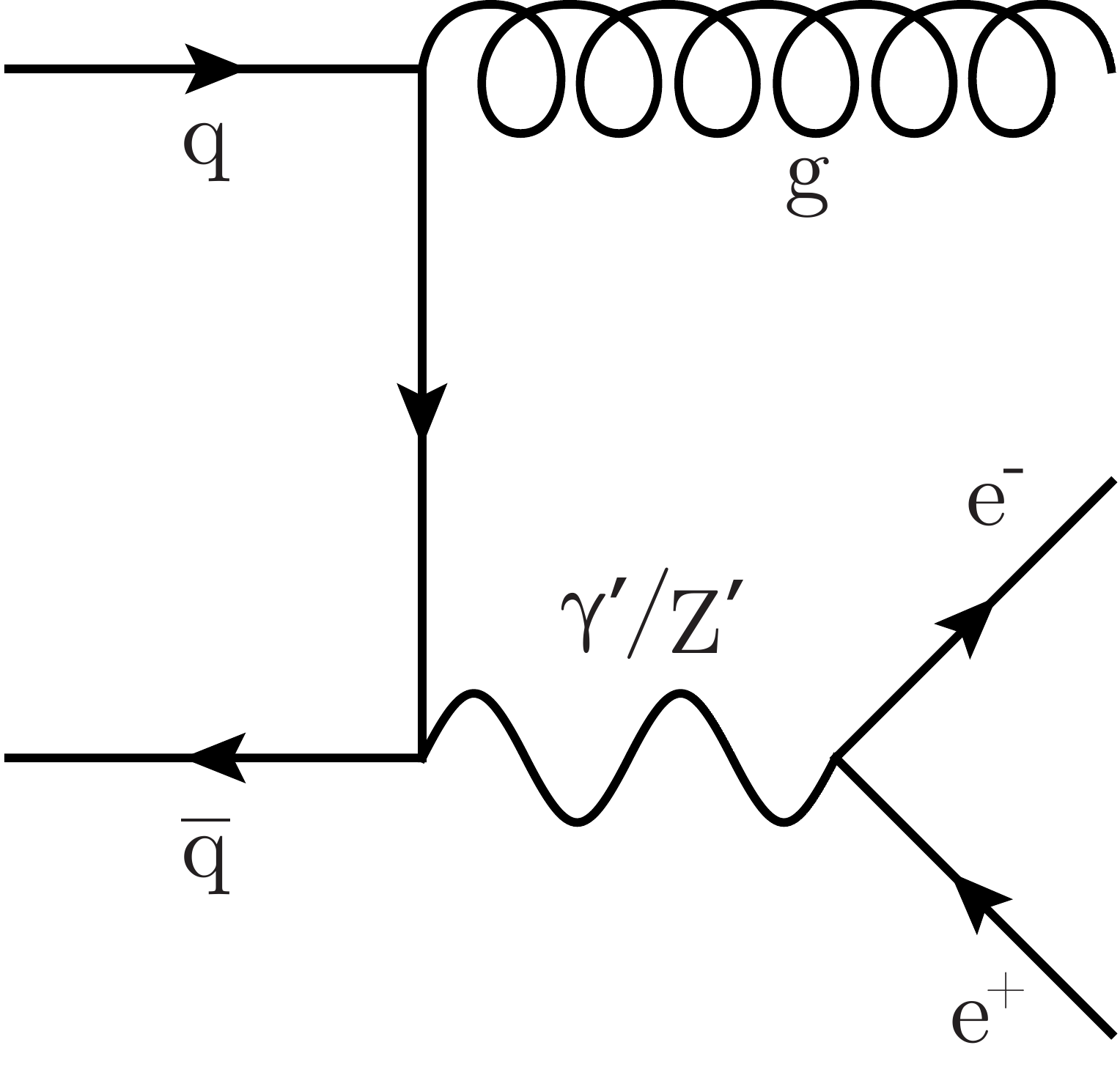}}
\caption{Feynman diagrams for the QGP processes $q \bar q \rightarrow e^+ e^-$, $q + g \rightarrow q + e^+ e^-$  and $q + \bar q \rightarrow g + e^+ e^-$. In each case the $e^+ e^-$ production is mediated by either the exchange of a virtual $\gamma^{\prime}$ or $Z^{\prime}$. We define $Q_i$ as the SM charge of species $i$ relative to the elementary charge unit $e$. For the $Z^{\prime}$, $V_i$ and $A_i$ are the SM vector and axial-vector couplings for species $i$ and $\tilde{g} = \frac{e}{2 \sin{\theta_W} \cos{\theta_W}}$. Note that t-channel and u-channel versions of the lower diagrams are also present.}
\label{fig:diagrams}
\end{figure}

Light (sub 10 GeV) dark gauge bosons are expected to contribute to dileptons production through the same processes as virtual photons. The Feynman diagrams for the dilepton production processes $q \bar q \rightarrow e^+ e^-$, $q + g \rightarrow q + e^+ e^-$  and $q + \bar q \rightarrow g + e^+ e^-$ are shown in fig. \ref{fig:diagrams}, mediated by either a $\gamma^{\prime}$ or $Z^{\prime}$.
The rate for such a process should be greatly enhanced when the invariant mass of the pair $m_{ee}$ is around the mass of the new gauge boson, due to the s-channel resonance (even if the couplings are suppressed).

Here we propose to exploit such a resonance to set limits on new, GeV-mass, dark gauge bosons. Before we proceed, it is worth considering whether such a signal could be detected in dilepton spectra from proton-proton collisions at GeV-scale, as well as from the QGP in heavy-ion collisions. The signal from Drell-Yan production of dileptons, used to set bounds for heavier gauge bosons \cite{Aad:2012hf,Chatrchyan:2012oaa}, is approximately an order of magnitude below the hadronic background for GeV-scale invariant masses \cite[sec. 4.1]{Dahms:2008bs}. Hence any enhancement due to the exchange of a dark gauge boson would be effectively invisible in prompt (proton-proton) collisions. The situation is different 
for heavy-ion collisions, since the QGP presents an additional thermal  source of dileptons for $m_{ee}$ of GeV-scale, which is much stronger than that from non-thermal prompt production \cite{PhysRevC.55.961,Rapp:1999zw,PhysRevC.63.054907}. This is why we focus only on thermal production from the QGP in this work and disregard the sub-dominant non-thermal production.

We will therefore search for an enhancement due to a $\gamma^{\prime}$ or $Z^{\prime}$ in the Au-Au dilepton spectrum for $1.2 \, \mathrm{GeV} < m_{ee} < 2.6 \, \mathrm{GeV}$, where the contribution from the QGP is expected to be largest, and competitive with the hadronic background. To calculate the dilepton spectrum for $\gamma^{\prime}$ or $Z^{\prime}$ we follow the same method as for virtual photons in sec. \ref{sec:dilepton_prod}, but replace the photon in the propagator by the dark gauge boson, as in the processes of fig. \ref{fig:diagrams}.

\begin{figure*}[t]
\centering
\subfloat{ \includegraphics[scale=0.38,trim=30 0 45 0]{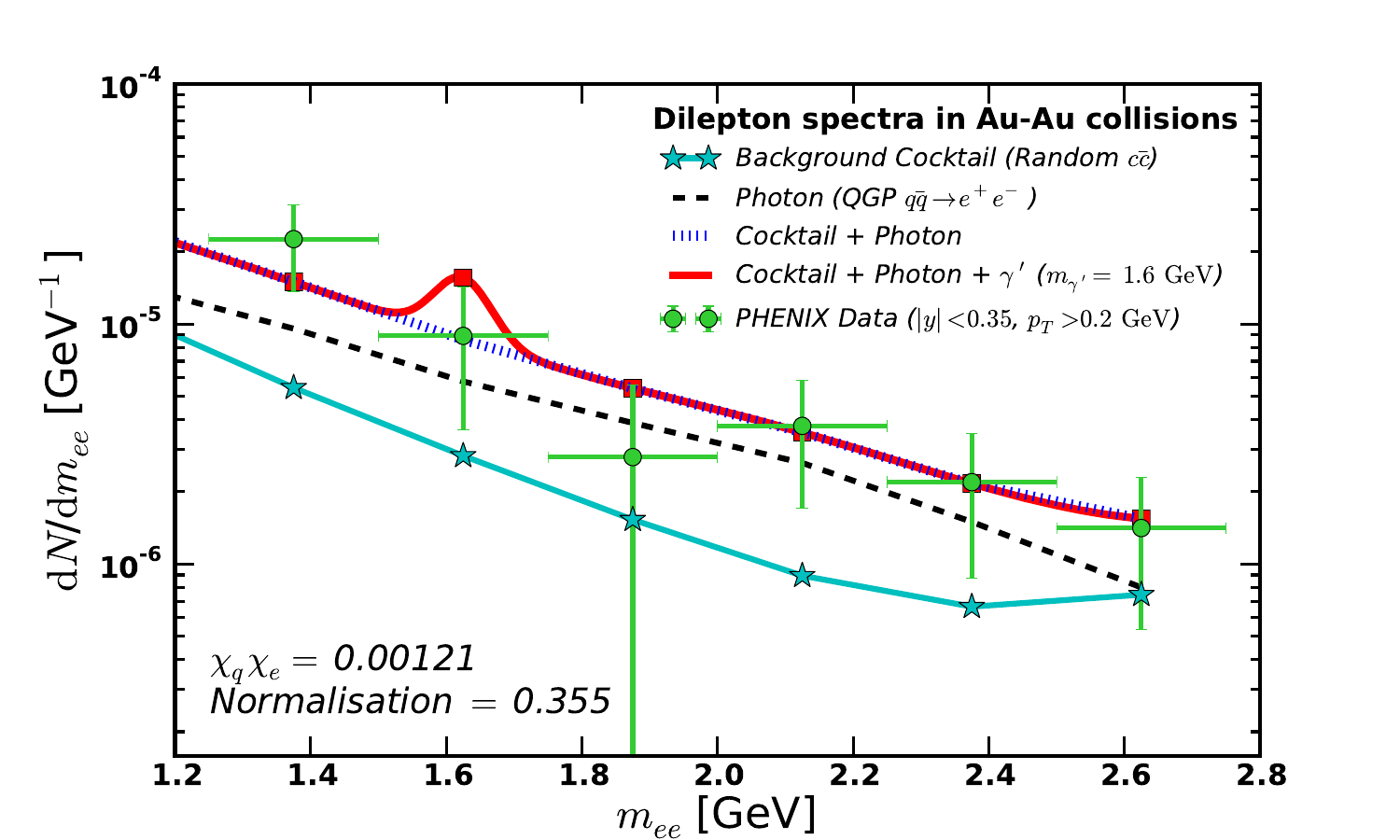} \label{m1600rnd}}
\subfloat{ \includegraphics[scale=0.38,trim=30 0 45 0]{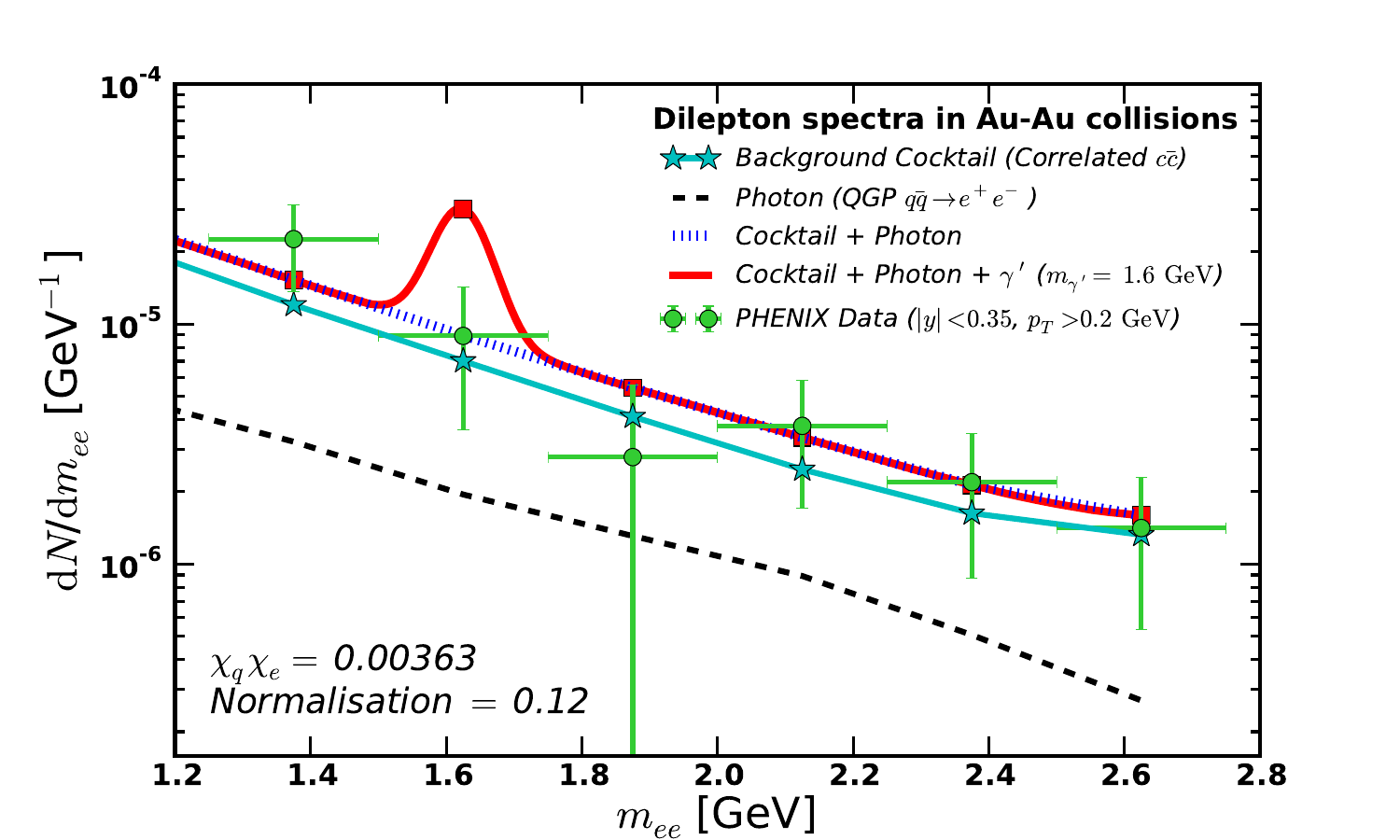} \label{m1600corr}}
\caption{Spectra of dileptons produced via $q \bar q \rightarrow e^+ e^-$, where the quarks exist as thermal partons in the QGP and the mediator is either a virtual photon or $\gamma^{\prime}$. This is compared with PHENIX heavy-ion data \cite{Adare:2009qk} and the hadronic background cocktail, dominated by either random $c \bar c$ (left) or correlated $c \bar c$ (right). The resonance from the virtual $\gamma^{\prime}$ is just visible due to its suppressed couplings to quarks and leptons. The photon and $\gamma^{\prime}$ spectra have been calculated at the measured values of $m_{ee}$ and binned in units of $\Delta m_{ee} = 0.25 \, \mathrm{GeV}$ (shown as $\color{red}{\blacksquare}$ when added to the background). Their normalisation has been allowed to vary, with the best-fit value used here. The lines are obtained by interpolating between these points, hence the width of the resonance is only an approximation to the true decay width to $e^+ e^-$, as discussed in the text. }
\label{fig:m1600_wphenix}
\end{figure*}

\vspace{-10pt}
\subsection{Dilepton backgrounds at PHENIX}
The background for dilepton emission, over the full possible invariant mass range, originates from various hadronic decays, referred to collectively as the ``cocktail''. In the IMR, there is some ambiguity in exactly how large the hadronic background is. 
One nevertheless expects the dominant background to be from $W$-boson decays of charm and anti-charm quarks; where the electrons and positrons are mistaken for dilepton pairs originating from a single vertex \cite{Lang:2013wya,Adare:2009qk,Mannien:2011b,PhysRevC.55.961}. 

In proton-proton collisions the production of $c$ and $\bar c$ quarks results in correlated decays, since they are themselves produced back-to-back from the same vertex. Hence the correlated opening angle of the detected $e^+$ and $e^-$ from the decaying $c$ and $\bar c$ is more likely to be close to $\pi$ than $0$, increasing the likelihood that they will be mistaken for a high invariant-mass pair. This results in a large dilepton background in the IMR, precisely where we hope to see a signal from the QGP in heavy-ion collisions. 

However for Au-Au there is evidence to indicate that $c$ and $\bar c$ scatter in the nuclear medium \cite{PhysRevLett.98.172301,Wang:1996yf}, which should effectively destroy such correlation, resulting in smaller opening angles on average and hence a softer $c \bar c$ background for nuclear collisions \cite{Lang:2013wya,Adare:2009qk,PhysRevC.87.014905,PhysRevLett.104.132301}. The first such scenario is referred to as  the ``cocktail" with correlated $c \bar{c}$ background, while the second is described as originating from random $c \bar c$ and is referred to as ``cocktail" plus random $c \bar{c}$. In principle the expected background is somewhere in between the two scenarios, depending on the degree to which $c \bar c$ scatter in the nuclear fireball. Hence both backgrounds are considered when setting limits in this work, similarly to the method of the PHENIX collaboration \cite{Adare:2009qk}.

\subsection{Signature of  the new gauge boson}

Shown in figure \ref{fig:m1600_wphenix} is an example of the dilepton spectra originating from thermal quark interactions in the QGP in presence of a new gauge boson ($f_{\gamma^{\prime}}(m_{ee},\chi)$, here taken to be a $\gamma'$ for the sake of the illustration with a mass of 1.6 GeV) and in the case of virtual photons only $f_{\mathrm{photon}}(m_{ee})$. Additionally the two hadronic background scenarios $f_{\mathrm{bg}}(m_{ee})$ are displayed, as mentioned above. The couplings in this figure have been chosen so that the contribution of the $\gamma'$ becomes visible above the photon signal and background. Note that only the  $q \bar q \rightarrow e^+ e^-$ process has been used here, since it is dominant in the invariant mass region considered, and the plasma has been assumed to be in equilibrium throughout its evolution. However the same resonance is present in all partonic spectra (e.g. $q + \bar q \rightarrow g + e^+ e^-$), and so our results are largely independent of the exact production process, provided perturbation theory holds.

The sum of these contributions ($f(\chi,N)$ in eq. \ref{eq:ftotal}) is represented by the red solid line in fig. \ref{fig:m1600_wphenix}. There should also in principle be a contribution from the hot hadron gas (HHG) i.e. dileptons from interactions between the mesons and baryons produced in the nuclear fireball \cite{PhysRevC.63.054907,Peitzmann:2001mz,Dusling:2009ej,Dusling2009212,Rapp:2013nxa}. The dilepton rate from the HHG should be subdominant to that from the QGP for the range of $m_{ee}$ considered here, and so is not incorporated into our analysis. The same is also assumed for prompt Drell-Yan production of dileptons \cite{Hwa:1985xg,PhysRevC.55.961,Rapp:1999zw,PhysRevC.63.054907,PhysRevC.87.014905}, produced when the nucleons collide before the plasma is formed. Note that these are additional potential sources of a dilepton enhancement due to an $\gamma^{\prime}$ or $Z^{\prime}$ and their inclusion would likely strengthen our derived limit\footnote{Indeed, although the prompt Drell-Yan contribution is smaller than the $c \bar c$ background in this invariant mass region, a limit could also be set in principle using this prompt signal. However such a limit would always be weaker than that set using the larger thermal yield from the QGP, or using both signals together.}.

The results are compared with the most recent Au-Au data from the PHENIX experiment  \cite{Adare:2009qk}. As one can see the main feature of the new gauge boson is an excess of dileptons, from thermal production in the QGP, at 1.6 GeV (for $m_{\gamma'}=1.6$ GeV) in the total spectrum, due to the resonance  in the s-channel production of the dilepton final state. Replacing the $\gamma^{\prime}$ with a $Z^{\prime}$ results in a similar resonance, hence it should be possible to set strong limits on the quark and lepton couplings, similarly to searches in proton-proton dilepton spectra for heavier gauge bosons.

One can draw a direct comparison between the resonance here, from the s-channel exchange of a new gauge boson in thermal dilepton production, and those from hadronic decays such as $\phi$ and $J/\psi$. The signature for either should be largely similar, however in our case the width of the resonance will depend on $\chi_q \chi_e$ and potentially also on a coupling to dark matter. One can obtain a first-order estimate of the width by requiring $\mathrm{d} N_{\gamma^{\prime}} / \mathrm{d} m_{ee} \ge \mathrm{d} N_{\mathrm{photon}} / \mathrm{d} m_{ee}$, since the photons constitute an irreducible background to the new gauge boson resonance. Following this method we obtain an approximation for the width of the $\gamma^{\prime}$ resonance\footnote{The formula for the $Z^{\prime}$ width is more complicated in principle, due to the potential axial-vector couplings which are absent for the photon, but the size should be similar to that of the $\gamma^{\prime}$.} $\Delta m$ to be,
\begin{equation}
\Delta m = m_{\gamma^{\prime}} \left(\frac{1}{\sqrt{1 - \sqrt{\chi_q \chi_e}}} - 1 \right).
\end{equation}
Hence assuming a value of $\chi_q \chi_e = 10^{-3}$, a negligible coupling to dark matter and $m_{\gamma^{\prime}} = 2 \, \mathrm{GeV}$ we obtain an approximate resonant width of $30 \, \mathrm{MeV}$. 
This is about an order of magnitude below the bin-size used in fig. \ref{fig:m1600_wphenix}, hence a more sensitive search using smaller bins should be eminently suitable to discover or set bounds on such a resonance.  
Adding a coupling to dark matter would change the estimate of the width and introduce invisible decay modes if $m_{\gamma',Z'} > 2 m_{DM}$.

Due to the uncertainties on the choice of the background, we have introduced a normalisation to estimate the QGP contribution. However we marginalise over it to set our limits, separately for either background scenario, as discussed in more detail in the next section. In figure \ref{fig:m1600_wphenix}, the normalisation factor for the photon and $\gamma^{\prime}$ signal has been chosen to be close to the value for which the fit between signal and data is best. 

 Comparing the two background scenarios in the fits of fig. \ref{fig:m1600_wphenix}, it appears that the dilepton signal from the QGP must be suppressed to fit the data when combined with the correlated background (as compared to the case of random $c \bar c$), and hence the enhancement from the virtual $\gamma^{\prime}$ is less visible. Hence if indeed the $c \bar c$ background is correlated as with proton-proton collisions, then the suppressed QGP emission should also result in weakened bounds on the $\gamma^{\prime}$ and $Z^{\prime}$ couplings. 

However for an uncorrelated charm-background the QGP emission provides a much larger contribution to the total spectrum. Hence there is a clear excess of the data above the uncorrelated $c \bar c$ background (in the IMR) which the QGP emission fills. One would therefore expect the bounds on the $\gamma^{\prime}$ or $Z^{\prime}$ resonance to be correspondingly stronger.

\subsection{Constraints on the new gauge boson couplings}
\label{sec:constraints}
As one can already see from fig. \ref{fig:m1600_wphenix} if modelling efforts for the QGP production of dileptons are indeed correct  \cite{Peitzmann:2001mz,Linnyk:2010vb,Linnyk:2011vx,Dusling:2009ej,Peshier:1994zf,Peshier:1995ty,PhysRevC.87.014905,Dusling2009212,PhysRevC.63.054907,Aurenche:2002pc,1126-6708-2002-12-006}, then bounds can be placed on the coupling of GeV-scale new gauge bosons to quarks and leptons.

For this purpose,  we shall define the limit by integrating under the normalised posterior volume, defined as $\mathcal{P} (f(\chi,N) | d) = \mathcal{L}(f(\chi,N) | d) \mathcal{P} (\chi) \mathcal{P} (N)$. Here $N$ is the normalisation of the signal defined above (common to both the photon and $\gamma^{\prime}$ signals)  and $\chi = \sqrt{\chi_q \chi_e}$. The latter two functions are the priors, which will be assumed to be linearly flat, and $\mathcal{L}$ is the likelihood. We use the following definitions,
\begin{eqnarray}
\mathcal{P}(N) &\in& [0,N_{\mathrm{max}}] \\
\mathcal{P}(\chi) &\in& [0,1] \\
\mathcal{L} &=& \exp \left[ - \sum_i \frac{(f_i (\chi,N) - d_i)^2}{\sigma_i^2}  \right] \\
f (\chi,N) &=& N \cdot (f_{\mathrm{photon}} +  f_{\gamma^{\prime}} (\chi)) + f_{\mathrm{bg}} \label{eq:ftotal},
\end{eqnarray}
where $i$ sums over the $m_{ee}$ bins used for the analysis and $\sigma_i$ is the uncertainty in each value of the data $d_i$. The functions $f_{\mathrm{bg}}$, $f_{\mathrm{photon}}$ and $f_{\gamma^{\prime}}$ are identical to those discussed in the previous section, with the latter incorporating also an interference term between virtual photons and $\gamma^{\prime}$. Since we claim no prior knowledge on the normalisation $N$, we should take the limit where $N_{\mathrm{max}} \rightarrow \infty$. However this would result in an improper prior which we can not use to set a limit. Hence we choose $N_{\mathrm{max}}$ to be finite, but significantly larger than any feasible normalisation for the QGP signal, such that its exact value has no effect on the final limit.

For the actual value of $N$ one has two options, both of which we consider: the first is to pick a value of $N$ and then set a limit by integrating under $\mathcal{P} (f(\chi,N) | d)$ with $N$ fixed at a value $N_0$. The second is to marginalise $\mathcal{P} (f(\chi,N) | d)$ over $N$, to obtain the probability distribution $\mathcal{P}(\chi | d)$, which we use to set a limit on $\chi$.

In the first case we are presented with several choices for $N_0$. Limits can be set using the value of normalisation for which the QGP dilepton signal fits the data from PHENIX best, as shown in figure \ref{fig:m1600_wphenix} (labelled as Scenario 1). As discussed earlier in this best-fit scenario, the signal from the QGP is suppressed for the correlated $c \bar c$ background, relative to that from random $c \bar c$.

However this is not the only possibility within this method: one can instead take a scenario where such a fit is not realised. For example as mentioned previously our calculations result in a dilepton signal larger overall than in a previous work \cite{Dusling:2009ej} (and the comparison to data in \cite{Adare:2009qk}). Hence we have also set limits on $\chi$ with $N_0$ such that our expected QGP signal is of the same size as in this work (Scenario 2). Of course we can also set $N_0 = 1$ for either background scenario, thereby assuming no alteration to our calculated spectrum in setting limits. 

It appears difficult to justify using any one value of $N_0$ to set a limit. To make sure that our limit is independent of the choice of $N_0$, we use instead the method of marginalisation over $N_0$ which allows one to set a limit while taking account of many different possible values of $N$ (Scenario 3)\footnote{We have taken the prior for $N$ to be flat, indicating that we have no prejudice as to its expected value. However with a more expert analysis into the variability of the spectrum with parameters such as $T_{\mathrm{max}}$, this could change. One could even extend this method and marginalise over the effect of uncertainties on both the shape and size of the dilepton spectrum from the QGP.}. In practice this means that any limit we set on $\chi$ will receive contributions from all values of $N$ within the range $[0,N_{\mathrm{max}}] $, weighted by the quality of the fit to the PHENIX data. In addition one can effectively treat $N$ as a proxy for uncertainties in for example the initial temperature $T_{\mathrm{max}}$ and formation-time of the plasma (although these could also affect the $m_{ee}$-dependence of the spectrum, for large deviations from our values), as well as the effect of cuts on the data. 
\begin{figure}[htb]
\centering
\vspace{10pt}
\includegraphics[scale=0.45]{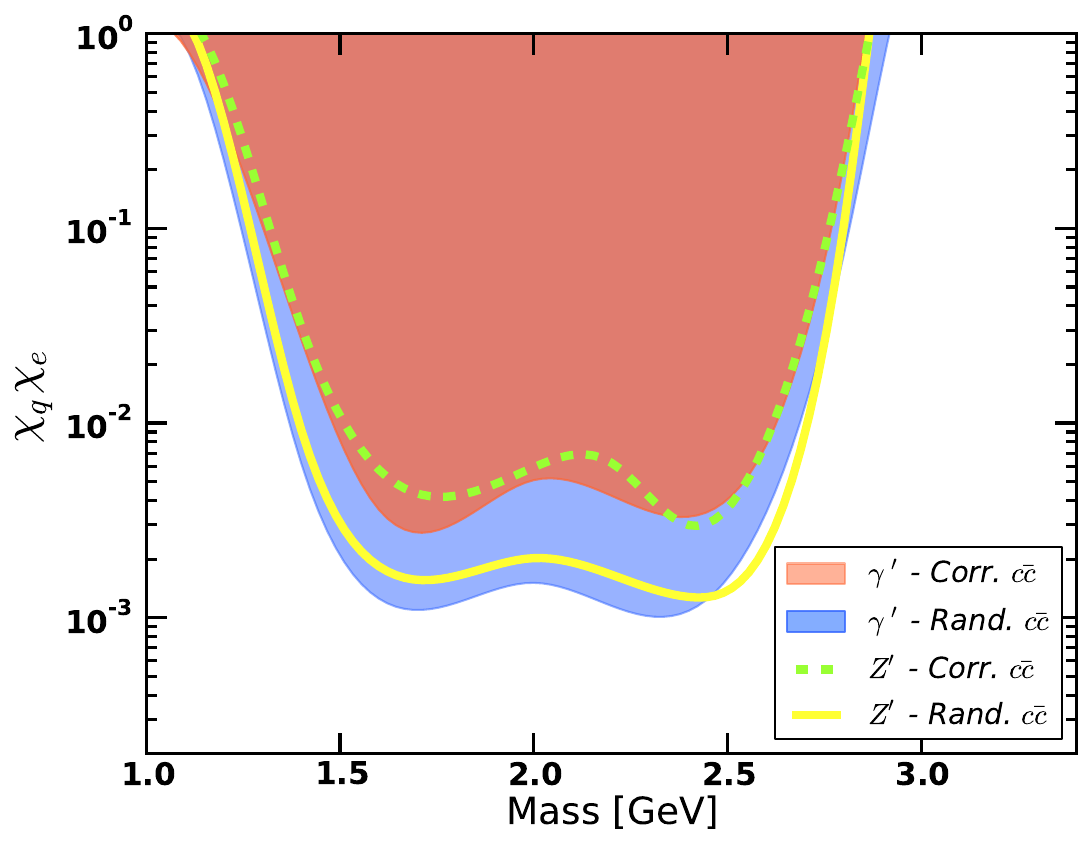}
\caption{Upper limits at $95 \%$ confidence on the coupling of a new gauge boson to quarks and leptons from the QGP dilepton signal in the IMR. For the $\gamma^{\prime}$ $\chi_q$ ($\chi_e$) is the relative coupling to quarks (charged leptons) as compared to the photon. For the $Z^{\prime}$, $\chi_q$ and $\chi_e$ are taken relative to the Standard Model $Z$-boson coupling to quarks and charged leptons. }
\label{fig:qqbar_limits_Aprime}
\end{figure}

The value of $\chi$ for which $95 \%$ of the volume of  $\mathcal{P}(\chi | d)$ (or $\mathcal{P} (f(\chi,N=N_0) | d)$ if we do not marginalise over $N$) is contained will define the limit for a given value of $m_{\gamma'}$, the mass of the $\gamma^{\prime}$ gauge boson. A similar procedure is also followed for a potential enhancement from virtual $Z^{\prime}$ exchange, with $f_{\gamma^{\prime}} (\chi)$ replaced by $f_{Z^{\prime}} (\chi)$. In this case we have taken $\chi_q$ as being the $Z^{\prime}$ coupling to quarks as a ratio to the coupling of the $Z$ (both vector and axial-vector), and similarly for leptons. Though there is no reason in general for the $Z^{\prime}$ axial and vector couplings to be related in the same way as for the $Z$.
 
By fitting such spectra to PHENIX data \cite{Adare:2009qk}, for a range of $\gamma^{\prime}$ and $Z^{\prime}$ masses, limits at $95 \%$ confidence have been derived assuming either a completely correlated or uncorrelated $c \bar c$ background for the dilepton signal. Shown in fig. \ref{fig:qqbar_limits_Aprime} are such exclusion bounds for the combined coupling of the new gauge bosons to quarks and leptons $\chi_q \chi_e$, for both background scenarios (and also marginalising over $N$).

Our strongest limit for the $\gamma^{\prime}$ corresponds to masses between $1.5 \, \mathrm{GeV}$ and $2.5 \, \mathrm{GeV}$ (which was to be expected given the invariant mass range used here). In this regime  $\chi_q \chi_e$ is forced to be smaller than $\sim 10^{-3}$. Hence if one assumes the most favourable scenario of a random $c \bar c$ background then such limits can be combined with those from purely leptonic experiments to bound the quark-$\gamma^{\prime}$ coupling $\chi_q$. As an example, taking $\chi_e \approx 2 \cdot 10^{-3}$ from the BaBar limits \cite{Aubert:2009au,Echenard:2012iq} one obtains $\chi_q \lesssim 0.5$ for $m_{\gamma'} \in [1.5,2.5]$ GeV.  For masses outside of this range the limit rapidly drops away, due to the potential enhancement being at the boundary of the IMR (for larger $m_{ee}$ the data are dominated by the $J/\psi$ peak and the QGP contribution becomes small).

It is important to study to what degree the limit changes if we do not marginalise over the normalisation, and instead employ one of the scenarios mentioned above, where $N$ is fixed at a value $N_0$. Limits under all such scenarios are displayed in the table below.

\begin{table}[h]\normalsize
\begin{center}
\begin{tabular}{  cc | c | }
\cline{2-3}
&  \multicolumn{1}{ |c| }{Rnd.}& Corr.  \\ \cline{1-3}
 \multicolumn{1}{ |c| }{Scenario 1 - Best-Fit} & $1.0  \cdot 10^{-3}$& $5.0  \cdot 10^{-3}$      \\ \cline{1-3}
  \multicolumn{1}{ |c| }{Scenario 2 - Suppressed} & $3.0 \cdot 10^{-3}$ & $5.0 \cdot 10^{-3}$      \\ \cline{1-3}
   \multicolumn{1}{ |c| }{Scenario 3 - Marginalised} & $1.5 \cdot 10^{-3}$ & $5.0 \cdot 10^{-3}$       \\ \cline{1-3}
\end{tabular}
\end{center}

\caption{Comparison of upper limits on $\chi_q \chi_e$ for the $\gamma^{\prime}$, derived under the various scenarios for the normalisation, as discussed above, for a new gauge boson of mass $1.6 \, \mathrm{GeV}$. For Scenario 1 we use the value of $N$ for which the QGP fits the data best when added to the background, and for Scenario 2 the QGP signal is suppressed to match that from \cite{Dusling:2009ej}. Scenario 3 is the limit in the case where $N$ is marginalised, as shown in fig. \ref{fig:qqbar_limits_Aprime}. We do not claim accuracy beyond $0.5 \cdot 10^{-3}$.}
\end{table}

For the $\gamma^{\prime}$, the weakest limit is in the case of the correlated background, for all scenarios. For a random $c \bar c$ background we see that Scenario 2 gives the weakest limit, since the signal from the QGP has been suppressed to match more closely the result from \cite{Dusling:2009ej}. However even with this suppression there is still a strong potential for the QGP to place bounds on the coupling of a $\gamma^{\prime}$ to quarks and leptons. The limit for the $Z^{\prime}$ behaves almost identically under each scenario.

In conclusion our preferred limit is that from Scenario 3 (fig. \ref{fig:qqbar_limits_Aprime}), where the normalisation has been marginalised over. However limits derived in the other scenarios are also valid, and do not deviate strongly from the marginalised bound.

Previous bounds on the coupling of the $\gamma^{\prime}$ in particular have generally taken $\chi_q = \chi_e = \chi$ \cite{Jaeckel:2013ija}, in which case our limit on the universal coupling $\chi$ is weaker than that from the BaBar experiment in the same mass range \cite{Reece:2009un,Aubert:2009au,Echenard:2012iq}. However although universal couplings are motivated by simple models for the $\gamma^{\prime}$, the validity of such a quark-lepton universality must still be tested. Hence our method, based fundamentally on quark (and gluon) interactions and dilepton production via a new gauge boson, can be seen as complementary to that from $e^+ e^-$ colliders such as BaBar, and should provide one with a test of new light gauge bosons without any specific assumptions about their characteristics (see e.g. \cite{Babu:1996vt,Fayet:2006xd}). Additionally if interactions of the new gauge boson are to help mitigate the tension between the Direct Detection experiments \cite{Agnese:2013rvf,Aprile:2012nq} a bound based purely on leptonic couplings, such as the one set using data from the BABAR experiment, has limited relevance compared to our result, where the quark-coupling is probed directly.

We note that results from simulations imply that correlations between $c$ and $\bar c$ are almost entirely lost \cite{PhysRevC.87.014905,Lang:2013wya} in nuclear collisions. Hence the (stronger) limit for a random $c \bar c$ background in fig. \ref{fig:qqbar_limits_Aprime} is likely to be more plausible.

\section{Prospects for future searches}
%=======================================
\label{sec:future}
As mentioned above the current precision results in an exclusion limit on $\chi_q$ for $\gamma^{\prime}$ which is only just smaller than unity, when combined with the latest bounds on $\chi_e$ from purely leptonic experiments. Ideally one would hope that with the increased sensitivity and centre-of-mass energy of future heavy-ion experiments (for example, the ALICE experiment at the LHC \cite{Koehler:2013mew}), the prospect of discovering a new gauge boson with couplings weaker than the bound set here would be eminently possible, provided they are not too small.

Alternatively if no discovery signal is seen, ALICE and other future experiments could improve the bound set in this work by several orders of magnitude at least (due in part to the stronger signal expected from the QGP \cite{Rapp:2013nxa}). Rather encouragingly, an observation of an excess of GeV-scale direct photons by ALICE has already been made \cite{Wilde:2012wc}, which is consistent with production from the QGP. With more precise data, the ability of the QGP to discover or set limits on new GeV gauge bosons should improve, especially if the bin-size of the data in $m_{ee}$ is reduced by an order of magnitude, which should make the $\gamma^{\prime}$ or $Z^{\prime}$ enhancement more prominent.

There is also cause for optimism from the QGP itself, since it is expected that the higher collision energy of nuclei at the LHC should result in the plasma being formed at a higher initial temperature and therefore lasting for longer before reaching $T_c$ \cite{Rapp:2013nxa}. One estimate for the initial temperature at the LHC is $T_{\mathrm{max}} \approx 500 \, \mathrm{MeV}$, compared with $\sim 400 \, \mathrm{MeV}$ for RHIC \cite{Dusling:2009ej}. As remarked upon earlier and shown in fig. \ref{fig:mee_spectra_photon}, the expected dilepton yield from the QGP depends strongly on $T_{\mathrm{max}}$ \cite{Peitzmann:2001mz}, and is several times larger for the potentially hotter plasma formed at the LHC, as compared to RHIC. Hence provided the background in the IMR does not also increase by the same factor, the hotter plasma produced in nuclear collisions at the LHC should provide an even stronger limit on the $\gamma^{\prime}$ or $Z^{\prime}$ coupling to quarks and leptons. The hope is that with a stronger signal, limits from the QGP will able to complement those from a future dedicated fixed target experiment \cite{Reece:2009un,PhysRevD.80.075018} for $1 \, \mathrm{GeV} \lesssim m_{\gamma/Z^{\prime}} \lesssim 2.6 \, \mathrm{GeV}$, as well as limits from parity-violation \cite{Bouchiat:2004sp}, meson/baryon \cite{PhysRevD.75.115004,Babusci:2012cr,Gninenko:2012eq,Beranek:2012ey} and heavy-quark \cite{Oh:2011nb} decays and proton-proton collisions at the LHC \cite{Jaeckel:2012yz,Aad:2012hf,Chatrchyan:2012oaa}.

\section{Conclusion}
\label{sec:conclusion}
By searching for an enhancement in the thermally-produced dilepton spectrum originating from the QGP in the invariant mass range $1.2 \, \mathrm{GeV} < m_{ee} < 2.6 \, \mathrm{GeV}$, we have bounded the product of the coupling of a new gauge boson to quarks  and leptons to be $\chi_q \chi_e \lesssim 10^{-3}$ at $95 \%$ confidence for a $\gamma^{\prime}$. Similar limits have also been derived for the $Z^{\prime}$.  One very powerful aspect of this work is that not only does it probe a new region of the gauge boson parameter space, by alleviating the non-universal couplings assumption, but it also enables to constrain the couplings to quarks and leptons simultaneously.

Our bound was derived assuming that the dominant background from $c$ and $\bar c$ decays \cite{Adare:2009qk,Mannien:2011b,PhysRevC.55.961} was suppressed due to interactions in the nuclear fireball, which destroyed any correlation between $c \bar c$ produced in the same interaction \cite{Adare:2009qk,PhysRevC.87.014905}. Although the case for such interactions is compelling \cite{PhysRevLett.98.172301}, weaker limits can still be derived in the case of a correlated $c \bar c$ background. As such it is possible to consider the correlated $c \bar c$ case as the most conservative limit  set in this work, especially in the case where $N$ is marginalised over also to mitigate the effect of uncertainties in the signal size. It would thus be difficult to justify setting a limit weaker than this with current PHENIX data \cite{Adare:2009qk}.

The dilepton spectra, for virtual photon, $\gamma^{\prime}$ and $Z^{\prime}$ exchange, were calculated within perturbation theory at leading-order, modified to include thermal masses for quarks and gluons due to a resummation of their propagators in the thermal medium \cite{Braaten:1989mz}. Although this is expected to work well for the dilepton masses considered in this work, it is still to some extent an approximation and constitutes a source of uncertainty to the derived limits. Contributions to the dilepton rate from additional processes such as multiple scattering \cite{Aurenche:2002pc,1126-6708-2002-12-006,Linnyk:2010vb} were included using code from \cite{Geliscode}. The effect on the new gauge boson resonance remains to be studied. The modification of the thermal QGP dilepton signal due to non-equilibrium effects was also studied; the rate of $q + g \rightarrow q + e^+ e^-$ is enhanced relative to the other processes, though not substantially. For the plasma at the LHC these processes may perhaps be competitive with $q \bar q \rightarrow e^+ e^-$. However in such a scenario the resonance due to new gauge bosons would still be present.

Further sources of uncertainty arise from ambiguity in the initial temperature of the QGP \cite{Peitzmann:2001mz,Peressounko:1999fs} and additional sources of dileptons such as Drell-Yan production \cite{Linnyk:2004mt} and the hadronic gas  \cite{PhysRevC.63.054907,Peitzmann:2001mz,Dusling:2009ej,Dusling2009212}. To an extent some of this uncertainty, especially in the initial temperature, is accounted for by marginalising over the normalisation of the photon and $\gamma^{\prime} / Z^{\prime}$ signal. Although such extra sources of dileptons should be sub-dominant to the QGP production in the IMR, their contribution should be included in a more precise analysis, and would likely enhance the resonance associated with the new gauge bosons. This in turn would result in stronger limits being derived.

Despite such uncertainties, we have shown that by exploiting the thermal dilepton signal from the QGP formed in heavy ion collisions, it is possible to set limits on the coupling of new gauge bosons to both leptons and quarks, at energy scales difficult to probe with previous collider searches. This is due to the stronger signal from thermal QGP radiation for invariant masses $1.2 \, \mathrm{GeV} < m_{ee} < 2.6 \, \mathrm{GeV}$, which is at least an order of magnitude larger than the non-thermal prompt signal used in previous new gauge boson collider searches.

Of course such bounds rely upon the existence of such a dilepton signal, however there is an abundance of evidence \cite{Adare:2009qk,Arnaldi:2009}  and theoretical models to indicate this is a fair assumption \cite{Peitzmann:2001mz,Linnyk:2010vb,Linnyk:2011vx,Dusling:2009ej,Peshier:1994zf,Peshier:1995ty,PhysRevC.87.014905,Dusling2009212,PhysRevC.63.054907,1126-6708-2002-12-006}. With upcoming data from the ALICE experiment \cite{Koehler:2013mew}, there is the very real prospect of detecting a new gauge boson with a mass of GeV scale, or else setting strong limits on its couplings to quarks and leptons, especially considering the hotter QGP predicted to form at the LHC \cite{Rapp:2013nxa}. Additionally, we chose to search for a resonance only in the IMR, due to the large expected QGP contribution and smooth background, however there is no reason why this could not be extended to lower or higher masses for a future study. There is perhaps potential even for the QGP to provide the means to probe other new physics scenarios beyond new gauge bosons \cite{Ellis:1990br}.

\section*{Acknowledgments}
We are grateful to F. Arleo, P. Aurenche and D.D'Enterria for their invaluable comments during the preparation stage of this manuscript. We would like to thank in particular F. Arleo with who CB initiated early discussions. JHD is supported by the STFC.

\end{document}